\documentstyle[12pt,here,epsf,epsfig]{article} 
\begin{document}
\setlength{\baselineskip} {2.5ex}
\begin{center}
{Contribution to Advanced Study Institute, 
"Symmetries and Spin" - Praha-SPIN-2002,\\
Workshop Chairman, M. Finger,
http://mfinger.home.cern.ch/mfinger/praha2002/
\\ Prague, Czech Republic, July 2002}\\
\vspace{0.4cm} 
{\Large\bf   First Observation of Doubly Charmed Baryons\\}
\vspace{0.4cm} 
{The SELEX Collaboration} \\ \mbox{~~~}\\
M.A.~Moinester$^{l}$,
G.~Alkhazov$^{k}$,
A.G.~Atamantchouk$^{k}$$^,$\footnotemark[ 1],
M.Y.~Balatz$^{h}$$^,$\footnotemark[ 1],
N.F.~Bondar$^{k}$,
P.S.~Cooper$^{e}$,
L.J.~Dauwe$^{q}$,
G.V.~Davidenko$^{h}$,
U.~Dersch$^{i}$$^,$\footnotemark[ 2],
A.G.~Dolgolenko$^{h}$
% ...
\\ \mbox{~~~}\\
% ...
$^e$Fermi National Accelerator Laboratory, Batavia, IL 60510, U.S.A.\\
% ...
$^h$Institute of Theoretical and Experimental Physics, Moscow, Russia\\
$^i$Max-Planck-Institut f\"ur Kernphysik, 69117 Heidelberg, Germany\\
% ...
$^k$Petersburg Nuclear Physics Institute, St. Petersburg, Russia\\
$^l$Tel Aviv University, 69978 Ramat Aviv, Israel\\
% ...
$^q$University of Michigan-Flint, Flint, MI 48502, U.S.A.\\
% ...
\vfill \rightline{\today} \newpage
\footnotetext[ 1]{deceased}
\footnotetext[ 2]{Present address: Infinion, M\"unchen, Germany}
\footnotetext[ 3]{Present address: Imperial College, London SW7 2BZ, U.K.}
\footnotetext[ 4]{Present address: Instituto de F\'{\i}sica da Universidade Estadual de Campinas, UNICAMP, SP, Brazil}
\footnotetext[ 5]{Present address: Kafkas University, Kars, Turkey}
\footnotetext[ 6]{Present address: Physik-Department, Technische Universit\"at M\"unchen, 85748 Garching, Germany}
\footnotetext[ 7]{Present address: The Boston Consulting Group, M\"unchen, Germany}
\footnotetext[ 8]{Present address: Instituto de F\'{\i}sica Te\'orica da Universidade Estadual Paulista, S\~ao Paulo, Brazil}
\footnotetext[ 9]{Present address: Lucent Technologies, Naperville, IL}
\footnotetext[10]{Present address: SPSS Inc., Chicago, IL}
\footnotetext[11]{Present address: University of Alabama at Birmingham, Birmingham, AL 35294}
\footnotetext[12]{Present address: Imadent Ltd.,\ Rehovot 76702, Israel}
\footnotetext[13]{Present address: S\"uleyman Demirel Universitesi, Isparta, Turkey}
\footnotetext[14]{Present address: DOE, Germantown, MD}
\footnotetext[15]{Present address: Solidum, Ottawa, Ontario, Canada}
\footnotetext[16]{ Present address: Siemens Medizintechnik, Erlangen, Germany}
\footnotetext[17]{Present address: Deutsche Bank AG, Eschborn, Germany}
\end{center}
 
\section {Abstract}

   The SELEX experiment (E781) at Fermilab has observed two statistically
compelling high mass states near 3.6 GeV/$\rm{c}^2$, decaying to
$\Lambda_c^+ \rm{K}^- \pi^+$ and $\Lambda_c^+ \rm{K}^- \pi^+ \pi^+$.  
These final states are Cabibbo-allowed decay modes of doubly charmed
baryons $\Xi_{cc}^+$ and $\Xi_{cc}^{++}$, respectively.  The masses are in
the range expected from theoretical considerations, but the spectroscopy is
surprising. SELEX also has weaker preliminary evidence for a
state near 3.8 GeV/$\rm{c}^2$, a high mass state decaying to $\Lambda_c^+
\rm{K}^- \pi^+ \pi^+$, possibly an excited $\Xi_{cc}^{++}$ (ccu*).  Data
are presented and discussed.

\section{Introduction}

\indent 
~~~~~~~The quantum chromodynamics hadron spectrum includes doubly charmed
baryons (DCBs): $\Xi_{cc}^{+}$ (ccd), $\Xi_{cc}^{++} (ccu), $ and
$\Omega_{cc}^{+}$ (ccs), as well as the triply charmed $\Omega_{ccc}^{++}$
(ccc).  A 1996 DCB review \cite{Moines} collected information on masses,
lifetimes, internal structure, production cross sections, decay modes,
branching ratios, yields, and experimental requirements for optimizing the
signal and minimizing the backgrounds. DCB works published since then are
given in Refs.  \cite {kis,ebert,gunter,oni,Gub,Gunt,Itoh, BPhys,nt}.  
The
doubly and triply charmed baryons provide a new window for understanding
the structure of all baryons.  As pointed out by Bjorken \cite{bj}, one
should strive to study the triply charmed (ccc) baryon. Its excitation
spectrum, including several narrow levels above the ground state, should be
closer to the perturbative regime than is the case for the DCBs. The (ccq)
studies are a valuable prelude to such (ccc) efforts.

\indent 
Hadron structures with size scales much less than 1/$\Lambda_{qcd}$
should be well described by perturbative QCD. The tightly bound color
antitriplet (cc)$_{\bar{3}}$ diquark in (ccq)  may satisfy this condition.
But the DCB radius may be large, if it is dominated by the low mass q
orbiting the tightly bound (cc) pair. The study of such configurations and
their weak decays can help to set constraints on models of quark-quark
forces \cite{fr,ros}.  Stong \cite {stong} emphasized how the QQq
excitation spectra can be used to phenomenologically determine the QQ
potential, to complement the approach taken for $Q\bar{Q}$ quarkonium
interactions.

\indent 
Savage and Wise \cite {sw} discussed the (ccq) excitation spectrum
for the q degree of freedom (with the (cc) in its ground state) via the
analogy to the spectrum of $\bar{Q}q$ mesons, where the (cc) pair plays the
role of the heavy $\bar{Q}$ antiquark.  Fleck and Richard \cite {fr}
calculated excitation spectra and other properties of (ccq) baryons for a
variety of potential and bag models, which describe successfully known
hadrons.  In contrast to heavy mesons, the descriptions of light quark
(qqq) and singly charmed (cqq) baryons are less successful.  We need to
better understand how the proton and other baryons are built from quarks.
The investigation of the (ccq) system should help put constraints on baryon
models, including light quark (qqq) and singly charmed (cqq)  baryons,
since the (ccq) has a quark structure intermediate between (qqq) proton and
$\bar{Q}q$ meson structures.

\indent 
In the double-charm system, there have been many predictions for
the masses of the J=1/2 states and the J=3/2 hyperfine excitations
\cite{BPhys}.  Most results are consistent with expectations of a ground
state mean mass around 3.6 GeV/$\rm{c}^2$.  The (cc)  color antitriplet
diquark has spin S=1. The spin of the third quark is either parallel
(J=3/2) or anti-parallel (J=1/2) to the diquark.  For (ccq), the J=1/2
states are expected to be lower than the J=3/2 states by around 80
MeV/$\rm{c}^2$ \cite {BPhys,nt,fr,lich}.

\indent 
Bjorken \cite {bj} and also Fleck and Richard \cite {fr} suggest
that internal W exchange diagrams in the $\Xi_{cc}^{+}$ decay could
reduce its lifetime to around $~100fs$, roughly half the lifetime of the
$\Lambda_{c}^{+}$.  Considering possible constructive interference
between the W-exchange and two c-quark decay amplitudes, it is possible
that this state should have an even shorter lifetime.

 We describe qualitatively the perturbative production mechanism for DCBs.  
One must produce two c quarks (and associated antiquarks), and they must
join to a tightly bound, small size anti-triplet pair. The pair then joins
a light quark to produce the final (ccq).  The two c-quarks may be produced
(initial state) with a range of separations and relative momenta (up to say
tens of GeV/c). In the final state, if they are tightly bound in a small
size (cc) pair, they should have relative momentum lower than roughly 1
GeV/c. The overlap integral between initial and final states determines the
probability for the (cc)-q fusion process.  Such cross sections may be
smaller by as much as 10$^{-2}$-10$^{-3}$ compared to single-charm
production.  Aoki et al. \cite {aok} reported a low statistics measurement
at $\sqrt{s}$ =26 GeV/$\rm{c}^2$ for the ratio of double to single open
charm pair production, of $10^{-2}$. This $D\bar{D}D\bar{D}$ to $D\bar{D}$
cross section ratio was for all central and diffractive events. This high
ratio is encouraging for (ccq) searches.  Cross section guestimates are
given in Ref. \cite{Moines}.

\indent 
Brodsky and Vogt \cite {bro} suggested that there may be significant
intrinsic charm (IC) $c\bar{c}$ components in hadron wave functions, and
therefore also $cc\bar{c}\bar{c}$ components.  The double intrinsic charm
component can lead to (ccq) production, as the (cc) pairs pre-exist in the
incident hadron.  Intrinsic charm (ccq) production, with its expected high
X$_f$ distribution, would therefore be especially attractive. When a double
charm IC state is freed in a soft collision, the charm quarks should also
have approximately the same velocity as the valence quark. Thus,
coalescence into a (ccq) state is likely.  Cross section guestimates are
given in Ref. \cite{Moines}.

\indent 
The semileptonic and nonleptonic branching ratios of (ccq) baryons were
estimated by Bjorken \cite {bj} in 1986. He uses a statistical approach to
assign probabilities to different decay modes. He first considers the most
significant particles in a decay, those that carry baryon or strangeness
number. Pions are then added according to a Poisson distribution. The
Bjorken method and other approaches for charm baryon decay modes are
described by Klein \cite {kle}.  For the $\Xi_{cc}^{++}$, Bjorken \cite
{bj} estimated the $\Lambda_c^{+} \pi^{+} K^{-} \pi^{+}$ final state to
have 5\% branching ratio; while for the $\Xi_{cc}^{+}$, he estimated the
$\Lambda_c^{+} \pi^{+} K^{-}$ final state to have 3\% branching ratio. One
expects \cite {Moines} that roughly 80\% of the (ccq)  decays are hadronic,
with as many as one-third of these leading to final states with all charged
hadrons.

\section{Observation of DCBs at Fermilab SELEX} 

\indent 
The SELEX experiment (E781) at Fermilab \cite
{SELEX,RICH,lclife,lcprod} carries out DCB data analysis based on the
sample of 1630 $\Lambda_c^+$ and 10210 D-meson events from lifetime studies
of these particles.  SELEX used 600 GeV/c beams of $\pi^-$, $\Sigma^-$, and
protons to produce charm particles in Carbon and Copper targets, detecting
them with a magnetic spectrometer with high mass resolution ($\approx 5
~MeV/c^2$) and high proper lifetime resolution ($\approx$ 20 fs.), and
Cerenkov particle identification of particles above 25 GeV/c.

\indent 
SELEX has two statistically-compelling new high-mass states near
3.5 GeV/$\rm{c}^2$ decaying to $\Lambda_c^+ \rm{K}^- \pi^+$ (3460
MeV/$\rm{c}^2$)  and $\Lambda_c^+ \rm{K}^- \pi^+ \pi^+$ (3520
MeV/$\rm{c}^2$) decay channels of DCBs $\Xi_{cc}^{+}$ (ccd) and
$\Xi_{cc}^{++}$ (ccu)  \cite {mark,amst}. They appear to be members of the
DCB family of new particles: (ccd) DCBs composed of two charm quarks and
one down quark; and (ccu) DCBs composed of two charm quarks and one up
quark.  SELEX also has evidence \cite {Thesis} for a $\Xi_{cc}^{++}$ (ccu)
baryon near 3780 MeV/$\rm{c}^2$, a high mass state decaying to $\Lambda_c^+
\rm{K}^- \pi^+ \pi^+$, which may be a decay mode of an excited DCB
$\Xi_{cc}^{++}$ (ccu*).  The 3520 MeV/$\rm{c}^2$ state satisfies all
expectations for being a $\Xi_{cc}^{+}$ state.  Its mass is compatible with
most model calculations for the DCB ground state.  Its lifetime is shorter
than 30 fs at 90\% confidence.  The 3460 MeV/$\rm{c}^2$ state has the decay
characteristics of a $\Xi_{cc}^{++}$ state, and a comparable lifetime.  
However, it is difficult to understand the 60 MeV/$\rm{c}^2$ mass
difference between the Q=1 and Q=2 states if they are members of the ground
state isodoublet.  It is not yet confirmed that $\rm{J}$ = 1/2 for both
states.

\indent 
SELEX analysis and simulations continue for the charmed Lambda
($\Lambda_c$) decay-mode DCB data. SELEX also searches for the states
observed to date in decay channels where the final state has a charmed
D-meson rather than a $\Lambda_c$-baryon.  Continuing SELEX data analysis
aims to achieve a more complete understanding of the double charm sector,
production with different beam particles, decay modes, cross sections and
their A-dependence, X$_F$ and p$_t$ distributions.

\begin{figure}[tbc]
\centerline{\epsfig{file=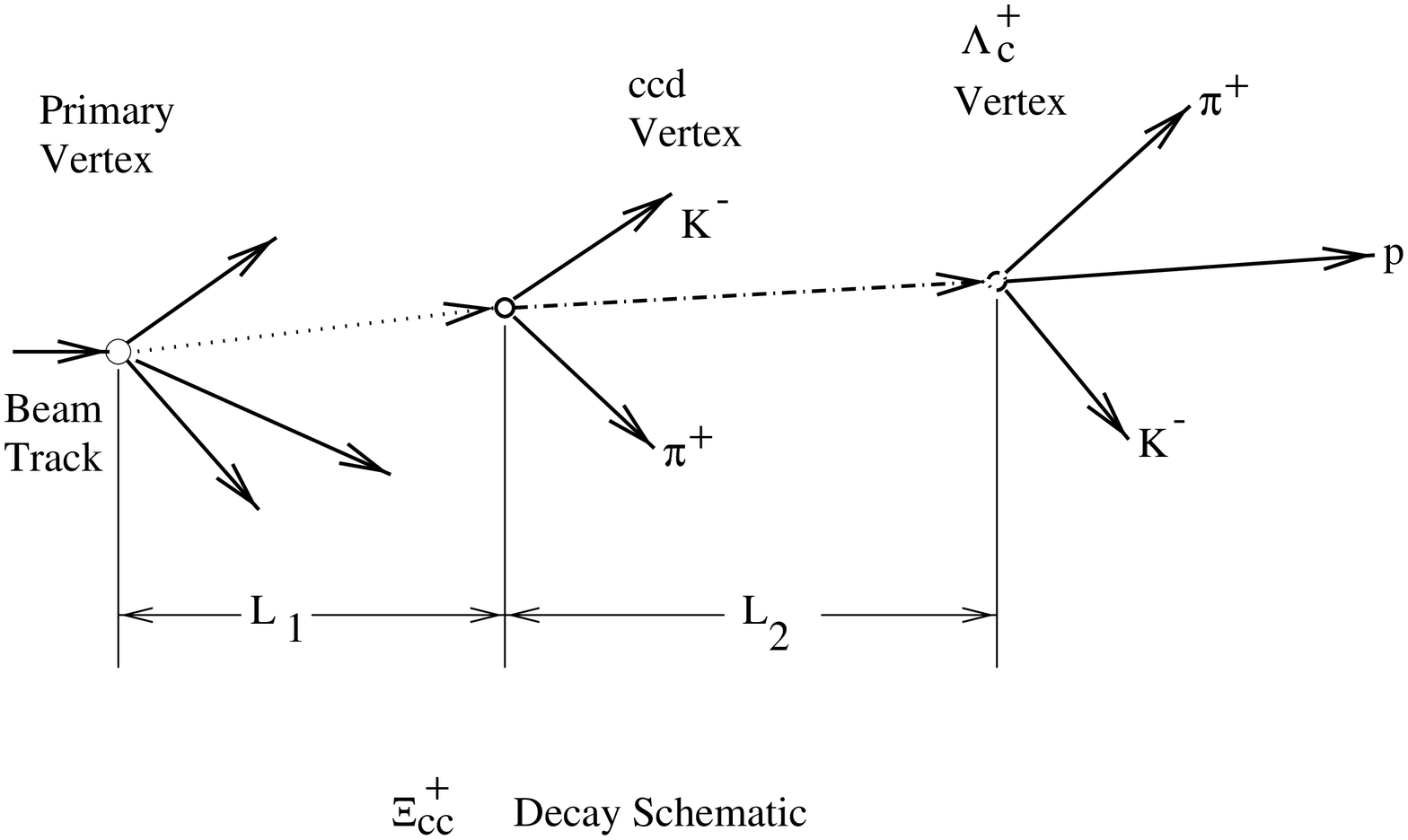,width=9cm,height=7cm}}
\caption{Schematic of $\Xi_{cc}^{+} \rightarrow \rm{K}^- \pi^+ \Lambda_c^+$} 
\label{fig:ccdfig}
\end{figure}

\subsection{Features of the Selex spectrometer}

\indent 
SELEX is a 3-stage magnetic spectrometer ~\cite{SELEX}.  The
negative 600 GeV/c Fermilab Hyperon Beam had about equal fluxes of $\pi^-$
and $\Sigma^-$.  The positive beam was 92\% protons. For charm momenta in a
range of 100-500 Gev/c, mass resolution is constant and primary (secondary)
vertex resolution is typically 270 (560) $\mu$m. A RICH detector labelled
all particles above 25 GeV/c, greatly reducing background in charm
analyses.~\cite{RICH}.  The details of single-charm analyses involving
$\Lambda _{c} ^{+} \rightarrow p K ^{-} \pi ^{+}$ reconstructions can be
found in ~\cite{lclife,lcprod}.

\subsection{Double-charm Analysis for $\Lambda_c$ Decay Modes}

\indent 
The double-charm search began with the sample of 1630 $\Lambda_c^+$
events (cud) used in the lifetime analysis. We ask for a weak-decay vertex
lying between the primary vertex and the observed $\Lambda_c^+$ decay
vertex.  A Cabibbo-allowed ccd decay can give a final-state $\Lambda_c^+$,
a $\rm{K}^-$, and a $\pi^+$.  The c decays weakly, for example by $ccd
\rightarrow csu + u\bar{u} + d\bar{d}$; corresponding for example to a 
$ccd \rightarrow \Lambda_c^+ \rm{K}^- \pi^+$ final state.  The event
topology contains two secondary vertices. In the first, a $\Lambda_c^+$ and
a $K^-\pi^+$ meson pair are produced. This vertex may be distinguished from
the primary vertex, if the (ccd) lifetime is sufficiently long. The
$\Lambda_c^+$ now propagates some distance, and decays at the next vertex.  
The experiment must identify the two secondary vertices. The idea is shown
schematically in Fig.~\ref{fig:ccdfig}.

\indent 
We reconstruct total electric charge states Q=1 and Q=2 in
separate reconstructions.  The Q=1 sample consists of two
oppositely-charged tracks along with a $\Lambda_c^+$.  The Q=2 sample
consists of a positively-charged triplet along with the $\Lambda_c^+$.  
Because the meson-pair tracks from the intermediate vertex may have
energies below the RICH detection threshold (25 GeV/c for pions), or since
they miss the RICH when they are emitted at wide angles, few achieve
particle identification in the RICH.  To be consistent with a $ccd
\rightarrow \Lambda_c^+ \rm{K}^- \pi^+$ final state, the negative track
should be a Kaon.  We build "right-sign" and "wrong-sign" samples, based
on calling the negative track a Kaon or a pion, respectively.

\indent 
Event selection cuts used here were taken without change from
previous single-charm studies.  For short-lived states,
$\rm{L_1}/\sigma_1 \ge 1$ and the $\Lambda_c^+$ momentum vector must
point back to the primary vertex within a $\chi^2$ cut.  We have
varied the cuts and observe that no signal significance depends
critically on any cut value.  The signals seen here are stable.

\begin{figure}[tbc]
\centerline{\epsfig{file=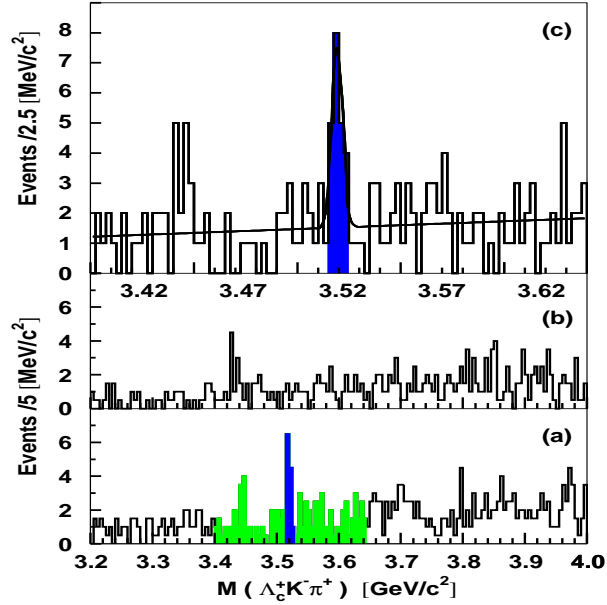,width=9cm,height=9cm}}
\caption{(a) The $\Lambda_c^+ K^- \pi^+$ mass distribution in 5 MeV/$c^2$ bins.
The shaded region 3.400-3.640 GeV/$\rm{c}^2$ contains the signal peak 
        and is shown in more detail in (c).
         (b) The wrong-sign combination  $\Lambda_c^+ K^+ \pi^-$ mass 
distribution in 5 
MeV/$c^2$ bins.
         (c) The signal (shaded) region (22 events) and sideband mass 
             regions (140 events) in 2.5 MeV/$\rm{c}^2$ bins.  
             The fit is a Gaussian plus linear background.}
\label{fig2}
\end{figure}

\subsubsection{Q = 1 reconstruction, 3520 MeV/$\rm{c}^2$}

\indent 
Fig.~\ref{fig2}.  shows the $ \rm{K}^- \pi^+ \Lambda_c^+$ mass
distribution for single-charged baryons. Fig.~\ref{fig2}(a) shows a
$\Xi_{cc}^{+}$ candidate at 3520 MeV/$\rm{c}^2$, consistent with most model
calculations.  The peak is 3 $\pm$ 1 MeV/$\rm{c}^2$ wide, narrower than
but consistent with simulation (5 $\pm$ .5 MeV/$\rm{c}^2$).  The final
state meets the double-charm Cabibbo-favored criteria: final state single
charm, final state baryon, final state $\rm{K}^-$.  The general agreement
between right-sign (negative track is Kaon)  and wrong-sign (negative track
is pion) average levels and fluctuations in Fig.~\ref{fig2} confirms that
most events are combinatoric background.  The signal channel shows a
22-event excess over a background of 6.1 $\pm$ 0.51 events, for a
single-bin significance of 6.3 $\sigma$.  Treating this with gaussian
statistics, the probability of such an excess is less than $10^{-8}$ for a
single bin.  We searched for a peak in the interval 3.2-4.3 GeV/$\rm{c}^2$,
or 110 bins.  In each bin there is a Poisson-distributed number of events
along with a Gaussian fluctuation on the background.  Summing these effects
gives the overall probability that our search would find such a fluctuation
anywhere in the search interval of less than $10^{-4}$.

\indent 
We calculated the lifetime likelihood for signal and sideband
regions in reduced proper time $ t^{*} ={M(L-L_{min})/ p c }$ for several
different lifetime values.  In this analysis $\rm{L_{min}} = \sigma_1$,
the error on the vertex separation $\rm{L_1}$ in the schematic from
Fig.~\ref{fig:ccdfig}.  From SELEX single-charm lifetime studies, we know
the lifetime resolution is about 20 fs.  Here, the sidebands show a
lifetime of $30_{-12}^{+18}$ fs.  The signal region result is shorter,
with lifetime less than 30 fs at 90\% confidence.  These two results are,
of course, statistically consistent with a common short lifetime for
signal and sideband regions.

\begin{figure}[tbc]
\centerline{\epsfig{file=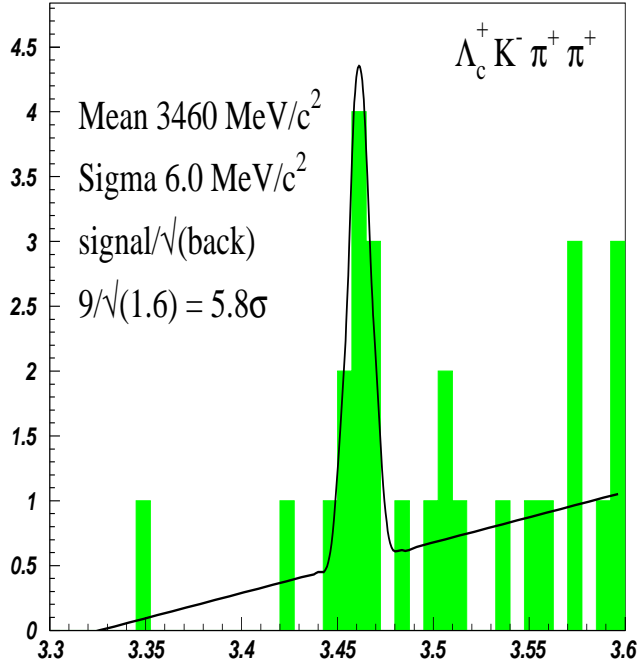,width=9cm,height=9cm}}
\caption{ $\Xi_{cc}^{++} \rightarrow \rm{K}^- \pi^+ \pi^+ \Lambda_c^+$  
mass distribution in 7.5 MeV/$\rm{c}^2$ bins.} 
\label{m6lo}
\end{figure}

\subsubsection{Q=2 Reconstruction, 3460 MeV/$\rm{c}^2$}

\indent 
Fig.~\ref{m6lo} shows a $\Xi_{cc}^{++}$ candidate at 3460
MeV/$\rm{c}^2$.  The peak is narrow, 6 $\pm$ 1 MeV/$\rm{c}^2$, matching
our simulation width.  The final state meets the double-charm
Cabibbo-favored criteria: final state single charm, final state baryon,
final state $\rm{K}^-$.  The Q=2 sample distribution is quite different
from that for Q=1.  The kinematic threshold for this channel opens near
3.0 GeV/$\rm{c}^2$, yet we have only 2 events below the 3460 mass peak.  
The background under the peak is clearly Poisson-distributed.  Making a
likelihood fit to a linear background function, we find 9 events in the
peak, compared to an expected background of 1.6 events.  The Poisson
probability that there is an excess of 7.4 events or more is 5 x
$10^{-5}$.  This search looked for an isospin partner to the
$\Xi_{cc}^+$(3520) and was limited to the range 3.3-3.6 GeV/$\rm{c}^2$,
containing 13 bins of width 22.5 MeV/$\rm{c}^2$.  The probability for a
statistical fluctuation up to a 7.4-event excess in any of these 13 bins
is less than 7 x $10^{-4}$.

\indent 
There are too few events to attempt a lifetime analysis, either
for signal or background.  We have compared the rest-frame angular decay
variables to a phase-space simulation.  The comparison is good for the
signal region.  The background tends to be more peaked along the flight
direction of the candidate, as one might expect for accidental
reconstructions.  

\indent 
One can ask why there are so few $\Xi_{cc}^{++}$ candidates
compared to $\Xi_{cc}^{+}$(3520) candidates.  Simulation shows that the
acceptance is substantially lower for the Q=2 system at masses below 3500
MeV/$\rm{c}^2$ than for a corresponding Q=1 system.  The SELEX
spectrometer cuts off the slow pion acceptance from the
higher-multiplicity state.  This can be seen in the data by the presence
of low-mass background in Fig.~\ref{fig2} and its absence in
Fig.~\ref{m6lo}.
 
\begin{figure}[tbc]
\centerline{\epsfig{file=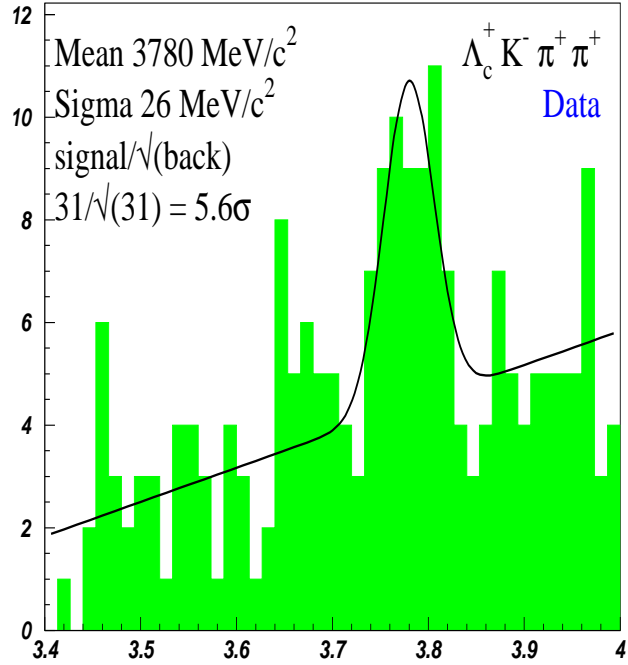,width=9cm,height=9cm}} \caption{
$\Xi_{cc}^{++} \rightarrow \rm{K}^- \pi^+ \pi^+ \Lambda_c^+$ mass
distribution in 7.5 MeV/$\rm{c}^2$ bins. 
The linear background is 
determined from a liklihood fit in the mass range 3.41-3.99 
GeV/$\rm{c}^2$.}

\label{m6fit_norm} \end{figure}

\subsubsection{Q=2 Reconstruction, 3780 MeV/$\rm{c}^2$}

\begin{figure}[tbc]
\centerline{\epsfig{file=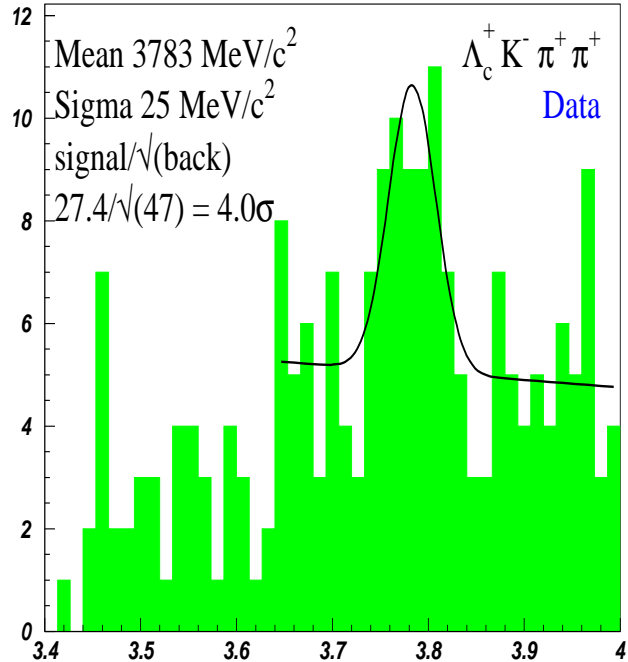,width=9cm,height=9cm}}
\caption{ $\Xi_{cc}^{++} \rightarrow \rm{K}^- \pi^+ \pi^+ \Lambda_c^+$  
mass distribution in 7.5 MeV/$\rm{c}^2$ bins.  The linear background is 
determined from a liklihood fit in the mass range 3.65-3.99 
GeV/$\rm{c}^2$.} 
\label{ccu_fit2}
\end{figure}

\indent 
We also searched at higher mass for $\Lambda_c^+ \rm{K}^- \pi^+ \pi^+$
decay of $\Xi_{cc}^{++}$ (ccu)  \cite {mark}.  We have evidence \cite
{Thesis} for a $\Xi_{cc}^{++}$ ccu baryon near 3780 MeV/$\rm{c}^2$, a
state which may be an excited DCB $\Xi_{cc}^{++}$ (ccu*).  The data are
shown in Fig.~\ref{m6fit_norm}, with a linear fit to the background.
Other fits are possible, with higher background levels, as in Fig.
~\ref{ccu_fit2}.  The background curves are determined from the spectrum
itself, from background simulation studies, and taking into account also
the wrong-sign mass plot.  The higher background in Fig. ~\ref{ccu_fit2}
is based on a simulation study, which shows that the excess near 3660 MeV
may be due to combinatorial background of a real ccd with a $\pi^+$ from
the primary vertex. This leads us to fit from 3650 MeV, to account for
this effect.  The number of signal events drops, the background increases, 
and the signal significance drops from 5.6$\sigma$ to 4.0$\sigma$.
The width of the peak is roughly three times wider than the
low lying DCB peaks, which would be possible for a state which decays
strongly.  For this ccu* however, the data do not have a plausible
explanation, in that roughly 50\% of the signal events above background
decay weakly (to $\Lambda_c^+ \rm{K}^- \pi^+ \pi^+$) and 50\% decay
strongly (to $\pi^+ \Xi_{cc}^{+}$). This is not possible for a single
state.  We can not now draw unambiguous conclusions.  We plan
further study of these data and backgrounds.

\subsection{Production}

\indent 
Production characteristics of the 22 signal plus background events
at 3520 MeV/$\rm{c}^2$ are indistinguishable from the single-charm
$\Lambda_c$ sample ~\cite{lcprod}.  The mean $p_t$ is 1 GeV/c and mean $x_F
\sim$ 0.33. This value of $x_F$ is significantly higher than the value $x_F
\approx 0.10$ that one expects \cite {Moines} if the (ccd) is produced near
threshold in a central collision.

We compared production of the $\Xi_{cc}^+$ and $\Xi_{cc}^{++}$
states by different beam hadrons.  The DCBs we see are produced solely by
the baryon beams in SELEX data. There are no signal candidates from the
pion beam.  We also checked if there is a dependence on the target nucleus.  
We found \cite {mark} ~ that the diamond/copper ratio of the signal events
is similar to the sideband events, which in turn behave like single-charm
production.
  
\indent 
Count rate estimates for DCBs at were given previously \cite
{Moines} based on ccq cross section of at most 25 nb per nucleon (25 nb/N).
This value is based on a reduction factor of 1000 per produced charm quark,
considering that the open charm production cross section (25$\mu$b/N) is
1000 times lower than the total inelastic cross section 25 mb/N. But SELEX
DCB cross sections are several orders of magnitude higher, as approximately
60 DCBs are observed in a sample of 1630 $\Lambda_c$.  Simulation studies
suggest that the double-charm states may account for as much as 40\% of the
$\Lambda_c^+$ sample seen in this experiment, a surprisingly high fraction.  
Work is in progress to calibrate the SELEX trigger efficiency factor, in
order to fix the absolute value of the DCB cross sections. The high DCB
cross section is reminiscent of the discovery of the $\Xi_c^+$ baryon in
the WA62 experiment at CERN, using a 135 GeV/c hyperon beam ~\cite{biagi}.
That cross section was also anomalously large and still has no good
theoretical explanation.  The FOCUS photoproduction experiment at Fermilab
has looked for these states using their $\Lambda_c^+$ events.  They see
very few candidate events and no signal peaks ~\cite{FOCUS}.

\begin{figure}[tbc]
\centerline{\epsfig{file=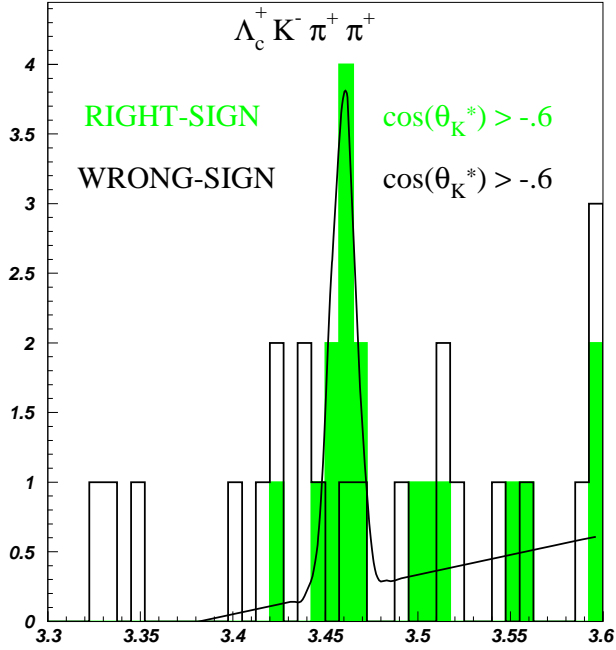,width=9cm,height=9cm}}
\caption{ $\Xi_{cc}^{++} \rightarrow \rm{K}^- \pi^+ \pi^+ \Lambda_c^+$  
mass distribution in 7.5 MeV/$\rm{c}^2$ bins, with angle cut.   
Signal events are shaded. Wrong-sign background is shown as open 
histogram boxes.} 
\label{fig:cculow69}
\end{figure}

\subsection{Background Reduction Studies for $\Lambda_c$ Decay Channels}

\indent 
The cleanest SELEX DCB data are for ccd and ccu lowest peaks,
while for ccu* the backgrounds are significant.  Consider the ccu at 3460
MeV.  Events outside the signal region show a strong preference for the
center-of-mass (CM)  angle of the negative track to be near 180 degrees,
where the angle is with respect to the $\Lambda_c^+$ direction.  This
suggests that these tracks represent accidental overlaps of a low energy
primary track and a real secondary vertex. In the CM system, a low energy
primary track is interpreted as a back-angle track from the secondary
vertex.

\indent 
Simulation indicates that a cut to remove such events should have
little effect on the signal region for a phase-space distribution. That is
indeed the case in the data.  For the ccu at 3460 MeV, with a cut
$cos(\theta_K^*) > -.6$, we obtain 1 background and 8 signal events.  The
spectrum is shown in Fig.~\ref{fig:cculow69}. This angle cut gives result
cleaner than without the angle cut, where we had previously 1.6 background
and 7.4 signal events. Without this cut, the Poisson probability that there
is an excess of 7.4 events in any of the 13 bins between 3.3-3.6
GeV/$\rm{c}^2$ is less than 7 x $10^{-4}$, and the signal significance is
5.8 $\sigma$.  With this cut, the Poisson probability that there is an
excess of 8 events in any of these 13 bins is reduced to $10^{-5}$, and the
signal significance is 7.9 $\sigma$. We will make similar studies for the
ccu* data.  The CM angle cut does affect the ccu* region strongly, but not
in a way that is simple to understand.  For ccu, ccu*, SELEX studies to
what extent the decay angular distribution of the $\rm{K}^-$ in the rest
frame of the $\rm{K}^- \pi^+ \Lambda_c^+$ system is compatible with a
phase-space simulation, both for signal and for background. With further
simulations, SELEX will try to better understand such experimental
distributions, and will explore other methods of reducing the backgrounds.

\subsection{Search for double-charm baryons in D-meson decay modes}

\indent 
SELEX started complementary DCB data analysis of these same states based
on our independent data for D-meson decays. Some possible decay modes of
interest are:  $(D^+ K^- p)$, $(D^0 \bar{K}^0 p)$ for (ccd);  $(D^+ \Lambda
\pi^+)$, $(D^0 \Lambda \pi^+ \pi^+)$, $(D^+ \bar{K}^0 p)$, $(D^{*+}
\Lambda \pi^+)$,  $(D^+ K^- p \pi^+)$, for (ccu). This work is in progress.

\section{Planning for the CERN COMPASS experiment}

\indent 
The SELEX efforts will help in the planning of a future state-of-the-art
double charm measurement via the CERN COMPASS experiment, and elsewhere. The
Sept. 2002 COMPASS-Future meeting at CERN included already a presentation of
SELEX data, and initial COMPASS plans in response to consultations with
SELEX on the SELEX DCB data \cite{lars}. Based on the SELEX results, one
could expect from COMPASS \cite {Moines,lars} up to 10,000 reconstructed
DCBs for a 100 day run. The SELEX data have stimulated more detailed
plans/prospects/efforts at COMPASS to study DCB and even triply
charmed baryon spectroscopy.         

\section{Summary}
 
We reviewed DCB theory.  We presented SELEX data for two
statistically-compelling new high-mass states that decay into a final state
$\Lambda_c^+$, $\rm{K}^-$ and one or two $\pi^+$, as expected for
double-charm baryon decays.  The 3520 MeV/$\rm{c}^2$ state satisfies all
expectations for being a $\Xi_{cc}^{+}$ state. Its mass is compatible with
most model calculations for the double-charm baryon ground state.  Its
lifetime is shorter than 30 fs at 90\% confidence.  The 3460 MeV/$\rm{c}^2$
state has the decay characteristics of a $\Xi_{cc}^{++}$ state.  However,
it is difficult to understand the 60 MeV/$\rm{c}^2$ mass difference between
the Q=1 and Q=2 states if they are members of the ground state isodoublet.
We have not yet tried to confirm that $\rm{J}$ = 1/2 for both states.  We
also showed evidence \cite {Thesis} for a $\Xi_{cc}^{++}$ (ccu) baryon near
3780 MeV/$\rm{c}^2$, a high mass state decaying to $\Lambda_c^+ \rm{K}^-
\pi^+ \pi^+$, which may be a decay mode of an excited DCB $\Xi_{cc}^{++}$
(ccu*). The SELEX data have been used already in the planning of the CERN
COMPASS DCB experiment. Internet links to SELEX and COMPASS DCB 
presentations are 
given in Ref. \cite {mm}. 

\section{Acknowledgements}
The authors are indebted to the
staff of Fermi National Accelerator Laboratory and for invaluable
technical support from the staffs of collaborating institutions.
This project was supported in part by Bundesministerium f\"ur
Bildung, Wissenschaft, Forschung und Technologie, Consejo Nacional
de Ciencia y Tecnolog\'{\i}a {\nobreak (CONACyT)}, Conselho
Nacional de Desenvolvimento Cient\'{\i}fico e Tecnol\'ogico, Fondo
de Apoyo a la Investigaci\'on (UASLP), Funda\c{c}\~ao de Amparo \`a
Pesquisa do Estado de S\~ao Paulo (FAPESP), the Israel Science
Foundation founded by the Israel Academy of Sciences and
Humanities, Istituto Nazionale di Fisica Nucleare (INFN), the
International Science Foundation (ISF), the National Science
Foundation (Phy \#9602178), NATO (grant CR6.941058-1360/94), the
Russian Academy of Science, the Russian Ministry of Science and
Technology, the Turkish Scientific and Technological Research Board
(T\"{U}B\.ITAK), the U.S. Department of Energy (DOE grant
DE-FG02-91ER40664 and DOE contract number DE-AC02-76CHO3000), and
the U.S.-Israel Binational Science Foundation (BSF).

\end{document}